\documentclass{article}
\usepackage{graphicx}
\begin{document}
\title{Interactions Among Agent Variables and Evolution of Social Clusters}
\author{Fariel Shafee\footnote{current address fshafee@alum.mit.edu}\\ Department of Physics\\ Princeton University\\
Princeton, NJ 08540\\ USA.}
\date{}
\maketitle
\section{abstract}
In this paper, we first review some basic concepts associated with a model for social interaction previously proposed by us.  Each agent is seen as an array of variables that can be found in different states.  The agents are then allowed to interact and form groups based on their variables.  We discuss how spin-glass type physics may be appropriate for our model. Several types of variables and costs associated with flipping the variables are discussed.  Then some simple graphs are presented to understand the formation of various levels of identities within social clusters.  In the end, we analyze events from the French revolution and the Russian revolution to to understand how different variables and identities interact within a hierarchical social structure.

\section{Introduction}	
Complexity in organization has lately been a field of intense study (Gellmann 2002; Wolfram 1995). Models of interacting systems have been applied to many real-world situations such as to predict market behavior (Black and Scholes 1973).

In social context too, interaction among agents is decisive in forming organizations.  Models treating each agent as a point in a graph have been used to study connectivity in social networks (eg. Newman et. al. 2001).  However, the relationship between agents in the social context is not simple.  Bonds can be formed and dissolved.  Although each agent is an individual, he is connected with others within more complicated structures.  In matter, larger structures of identity can often be reformed by interchanging a component with a similar one, but in the case of human societies, the complexity of each of the members make the dynamics more interesting, because the unique individual needs to be counted for within the context of an ensemble as well.

Previously (Shafee 2002; Shafee 2004; Shafee 2005), we have proposed a model regarding interacting heterogeneous agents, giving rise to organizations and notions of self at various levels.  In this paper, we review and summarize the basic concepts of our proposed model.  We then fit the concepts in terms of real world scenarios.  We observe how societies evolve and irrational behaviors emerge from interaction among variable states with time as the number of agents in a cluster changes.

\section{Interactions and Irrationality}

Conventional economics tries to predict the market and human behavior based on the assumption of rationally behaving agents dealing with one another in ways to maximize their own utility curves.  However, in an imperfect world, irrationality exists. Phenomena like war and spite contradict rational norms.

In conventional economics, a mathematical formulation involving equations for relevant variables is based on a set of needs for possible trades depending on supplies and demands of individuals. However, these clean relations yielding plots with distinct points of intersection displaying stable points or well-defined positions and payoffs fail to explain the so-called irrational states that often spread from one ill-defined point to large-scale social volatility, herd mechanism and fanaticism.  The difference of perceptions in individuals in individuals is well recognized (Hume, 1777) and may play a role.

Recently (Ostrom 1997), it has been shown that in complex organizations, an individual often makes decisions that are contradictory to optimal Nash equilibria.  The author also discussed how repeated games failed to teach the individual what the game theoretic optimal strategy is. Rather, increased communication often led to trust and caused the player to contribute more to the pool of commons.

In this paper, we try to formulate less than rational behavior in a more rigorous manner, which may lead to mathematical expressions for predicting social dynamics. We examine the fine structure of human perceptions and utilities and model possible chaotic or annealed behaviors of social lattices with coupled variables from a point of view of interaction physics, especially the spin glass model (Edwards Anderson 1975; Sherrington Kirkpatrick 1975). We introduce dynamics in the identities of interacting agents and take the matching components and contradictions  into account in these identities to find complicated behavioral patterns in interactions and dealings.

We assume that although agents can communicate to agree upon a broadly shared idea of the world, each agent is connected to her world independently, and gains information about the world by means of perceptions and interactions.  However, these agents are then again connected within social clusters.

In our model, the interactions among agents comprise a game of incomplete information, where each agent can guess only partial information about the other agents while calculating his next move. This ``transfer" of information about identity is achieved by means of repeated interactions. Hence, the information trapped in different agents may not always be transferable to another agent. Some of the pieces of information and preferences existing in different agents may be conflicting, inconsistent, and may at times be genetically determined and hence, almost permanently local to an individual agent. For example, low levels of latent inhibition to the environment may be associated with both the concept of genius or madness in two extreme cases (Peterson and Carson 2000). This suggests that some people are prone to be receptive to more stimuli from the environment and be more connected to the environment by default.

We argue that local rationality in the sense that each agent tries to maximize the needs of his own identity may in turn give rise to irrational behavior non-locally, and show how the grouping of various levels of identities may create volatility.

Previously, (Shafee 2002; Shafee 2004; Shafee 2006), various aspects of an interaction based identity model was discussed, and the possible effect of using the concept of changing identities in the example of a specific hierarchical social cluster was studied in (Shafee and Steingo 2007).  The constraints in the number of parameters expressed at a hierarchy level were discussed in (Shafee 2007) together with the concept of semi-closed identities. In this work, we start by comparing the notion of an identity-model based on interacting variables with models of identity existing prior to our propositions. Then we introduce some more detailed ideas and concepts.  Next, we formulate the components of the interaction Hamiltonian involving the variables that control the dynamics of the system. We review different types of variables (Shafee 2004) and study the effects of these variables at an individual and at a group level given interaction based potentials.  Finally, we study various social organizations in which agent and group identities evolve, because of the interactions between these different variables and observe how social organizations and dynamics originate due to interactions.

\section{Social Models Prior to Our Proposition}

Although existing literature has tried to find relatedness among agents (Pepper 2002, Hamilton  1964, Connolly and Martlew 1999), the models are often simplified as described below:

\subsection{Models based on Genetic Similarity}
Models exist to calculate match among genetically related agents (Dawkins 2006, Pepper 2000). However, recent mappings of the human genome have shown that human beings are genetically 99.9 percent similar (International Human Genome Sequencing Consortium 2001) and sometimes members of two isolated clusters may have more overall genetic matching than members of the same cluster [e.g. broad genetic difference in blood groups within the same group of people]. Despite this statistics, evidences of altruism and affiliation can be found within ethnic, social and cultural groups (e.g. see Bowles and Gintis 2004), and specific minuscule differences in genetic structure sometimes have large-scale expressions in interactions with the environment and other agents. Examples are: single mutations in one gene leading to deadly diseases like Huntington's disease (Imarisio et al 2008), and small mutation differences resulting in the broad expression of the type of melanocyte. Then again, a large percent of DNA remains as non-coding DNA or junk DNA in eukaryotic cells (Ohno 1972).

\subsection{Networks based on Homogenous Agents}
	The structures of social clusters and graphs have become a field of study recently (see e.g. White and Reichardt 2007; Gastner and Newman 2006;  Newman et al 2001; Barabasi and Reka 1999). However, detailed analysis of psychological and social causes leading to the fine structures of the clustering has not been performed.  The studies of social networks and the field of psychology and relatedness leading to clustering, therefore, remain disjoint.

	In some papers (Bowles 2004; Gintis 2003b), the so called group or pro-social emotions are mentioned, but the findings related to group emotions in non-genetically related clusters remain restricted to observation, and calculations of economic group stability assume that pro-social emotions are preprogrammed (Bowles 2004) or are parts of reciprocity between individuals (Bowles and Gintis 2003). However, no detailed explanation for these existing emotions was provided, though they were taken as traits that favored the existence of a group.
The dynamics of such emotions was not studied either.

\section{Encompassing Ideas into One Model}

These isolated and sometimes conflicting models can be incorporated into a larger model if some new observations are made and applied. Identity is not static. Extreme cases of the same person undergoing drastic personality changes can be observed after brain lesions or injury of the brain due to accidents (see e.g. Raitu 2004). Other physiological changes due to hormones etc. may also cause personality shifts  (Schulkin 1999). Mundane isolated interactions with an individual indicate time and experience gaps changing a personality, although a broad group of tags may remain constant, as well as shared memories. The ideas of continuity and changes within a person might be validated by his experiences, where the past rationally precedes the present, to update his preferences. However, to another person observing the first person with gaps of time and space between the two snapshots of the same person the change might not appear rational.

\subsection{Expression of Variables}
In (Shafee 2007), the expression of parameters at different levels of a complex system was explored.  Complexity and identity can be organized at various levels, where lower levels of complexity are contained within a higher level in a hierarchical fashion. So, the components of a complex system are contained within the system at lower levels of identity. For example, the components of an agent, such as  various organs are contained within him at a lower level so that the result of the interactions among the components is expressed at the higher human level, creating characteristics typical of human interactions.  Similarly, the complexity of a cluster derives from the organization and interaction of its component agents. Macroscopic average differences between clusters, and also finer differences among agents within these clusters indicate that there are some constraints and choices regarding how many variables can be expressed at each interaction level (Shafee 2007). Here the scale determines the nearest neighbors and the range of interactions.  So, each level will have a time scale for interactions as well as a space scale. The concept of scale with respect to complexity has been discussed in (Ahl and Allen 1996).

\subsection{Interconnectedness}
Different variables and states defining persons are sometimes complementary, and may also be mutually exclusive (contradictory), when placed together or allowed to interact. However, within each agent the variables that define an agent are connected together. Hence, when two agents interact, their matching and contradictory variables are introduced at the same time.

\subsection{Changing Priorities}
 The priorities or weights assigned to a preference or a piece of knowledge by an agent is important and sometimes often modifiable.  The introduction of a certain environment which is harmful may change an agent's priorities so that protection against that new environmental state must come first if the agent is to exist.  As the agent ages his biological parameters change and priorities are redefined as well.  Again, priorities vary from agent to agent depending on the agent's local environment and the agent's personal taste.  The introduction of risk factors in future possibilities may also cause agents to choose different priorities depending on whether the agent is risk prone or risk averse.  The degree of security and risk sought by an agent may also vary depending on personal and environmental situations and factors such as responsibility for other agents who might be genetic dependents.  As agents acquire more information and experience, priorities may change as well due to possible future predictions.

\subsection{Interactions and Updates}
Interactions among different types of traits have various degrees of stiffness and their interrelatedness often cause interesting dynamics (see Shafee 2004 for an example). The resistance to change preserves an identity, while change favors adapting to a new set of connections.  Adaptation is possible because of the existence of many possible suppressed states within the agent (for example, a large part of the genetic code in the human genome is not expressed.  Again, specific environments may trigger genetic expressions. see Gobbs 2003, Hasler 2007). The identity of an agent is derived from an idea of continuation so that modifications are made subject to some degree of stiffness and some constant states.

Interactions of agents with the environment provide them with information and the introduction of new environments may cause agents to modify some tastes or preferences as well (for example, changing the language of expression, food habit). Again, a sense of identity derived from interaction among agents may cause agents to modify their preferences.  A simple example is seasonal fashion where an agent modifies her preference for clothes to fit into a group by interacting with other agents.

\subsection{Level of Complexity and Fuzziness}
In a recent paper (Shafee 2009), the relation of the level of complexity with the fuzziness of the identity was discussed. At higher levels of complexity, the exact number of total components at the lowest levels is not defined since overlap and interactions among components create not a point but a band of stability.  For example, in a hadron, exactly three quarks form a more or less shielded, localized point of near stability.  However, in a larger human being, comprising many quarks, within many atoms that make up the organs connected together, the spatial extension and mass occur only within a viable range and not with a specific value.  The difference in shielding, and possible interactions with neighboring structures and a connected environment may also contribute to changing semi-stability as complexity level changes.

\subsection{Connection with the Environment}

Interactions of the agents with the environment may cause the inter-agent interactions to change as well.
In (Shafee 2002)  an axiom based game was constructed where an agent played against other agents and also the environment.  Both agents and the environment were provided with identities so that each player tried to align the other player along his most optimal direction. The vastness of the environment was matched with the number of agents playing together, and situations like flipping of axioms based on conversion were introduced.

The degree of match between two agents' axioms thus provides a measure of the total effort expended by agents in a coherent manner against the environment.  Hence, the number of agents sharing an axiom or preference gives a measure of that axiom's strength when playing against the environment. The axioms placed in a social network may be attributed an identity distributed among agents, and any bond among agents sharing an axiom would strengthen that axiom with respect to the environment.

\section{The Physics of Interaction Dynamics}

The analogy between spins and interacting variable (component) states in agents can be used to formulate detailed social dynamics.

A basic interaction potential using spin glass type interactions ( see Edwards Anderson 1975; Sherrington Kirkpatrick 1975) is given by:

\begin{equation}
H =  - J_{ij}^{ab}  s_i^a  s_j^b  -  h^a  s_i^a
\end{equation}

Here $J_{ij}^{ab}$ is the measure of coupling between attributes $a$ and $b$ between agents $i$ and $j$; $s_i^a$ and $s_i^b$ are the given variable states. $h^a$ is the component of the environment interacting with attribute $a$. The interaction Hamiltonian $H$ determines the net forces at play on each independent variable.

In (Shafee 2009), we will be modifying this basic spin glass equation by taking properties of social interactions into account and then form a master dynamical equation.  In this paper,  we explore our model qualitatively in terms of some specific social scenarios in order to validate the assumptions taken into account.

The use of a spin-glass type model is validated by noting that the system concerned is dynamic and is dependent on interactions among variables of different agents.  These variables can exist in different states, and interactions can flip the state of a variable in order to minimize the potential of the system consisting of many parts.  Similar models of complex systems depending on interactions between parts have been proposed (Hopfield 1975)

However, each agent is a complex entity with many variables.  Hence, the interaction Hamiltonian $H$ is somewhat similar to an anisotropic Heisenberg Hamiltonian in the sense that we have two arrays with different component values (variables in different states) interacting and the minimization of total energy in an energy landscape is the asymptotic goal. However, as the number of variable-states is increased, numerous local minima arise and fluctuations (noise) and sudden changes (including sharp changes in the environment and mutations) contribute to instability of the network in a local minimum. We call it an anisotropic Heisenberg model because the components of the vector do not have the same metric or weights in coupling together.

\section{Identity Units and Modified Utilities}
As described previously, each complex identity unit $i$ can be expressed as a localized coupled set of attributes $a$ (comprising many lower levels of attributes that may not all have a net expression).  Interactions among those coupled variables produce local interaction potential minima in the Hamiltonian function, and hence a degree of stability that is robust to some degree against outside interactions that tend to raise this local interaction potential.  The concept of identity is thus related to the continuation of local interaction minima existing on different scales, with robustness deriving from each local minimum  comprising  much smaller strongly coupled minima contained within the structure in a complex manner.
	
\subsection{Identity of an Agent}
Theories about how an agent perceives his own traits differently from the traits of others exist in current psychology literature (see Klein et al 2004, for example). Such theories justify our notion of interacting agents with variables in different states.

An agent's identity is also dependent on the persistence of a memory, which is a sequence of correlated events (see, for example, Raaijmakers 1979,  Raaijmakers and Shiffrin 1981) that are past interactions between the agent's existing states and the states of other agents or the environment. In human memory, certain events can produce cues (Padilla-Walker and Poole, 2002). Memories, thus, can be seen as change of  states (in neuron connections), such that a specific interaction can send a cue to stimulate the memory of another set of interactions or a changed state.  So memories can be seen as states within the agent that act as an interface between the agent's own states and the environment ( or other agents).

Hence, an agent's identity derives from the balance of being able to choose a future, given a locally preserved known past (see Palis 2002 for a discussion) in order to reach the most optimal interaction potential possible, and  the stiffness to change i.e. to preserve its current states and couplings.

However, the reactions evoked within an agent based on past interactions, which can be seen as part of the agent's identity, come in segments (Raaijmakers and Shiffrin 1981).  Instead of calculating long chains of consecutive events, short segments are correlated to evoke instantaneous reactions.  Repeated exposure to a sequence conditions a reflex (Pavlov 1927) which is a measure of correlated sequence clauses.  Hence, segments of correlated interactions become part of the agent as programmed instantaneous reactions, while calculating a large chain of events requires time and energy.  Stochastic measure of immediate interactions and retrieved specific long chains of events give rise to the agent's perception of short term and long term identity and utility.  Similarly constructing actions for short and long term futures that are optimal require predicting highly correlated sequences, and probabilistic segments in a chain, involving risk factors.

The current state space of an agent is a function of his genetic design and also his past interactions. So the uniqueness of the agent derives from the uniqueness of his local past and also the uniqueness in genetic make-up.  Even if two identical twins might have  exactly the  same genetic make-up, their uniqueness appears from separate local pasts that are stored within modified states.

\subsection{Identity of a Cluster}

An agent is part of a cluster of agents, and the identity of the cluster is created by taking into account the dynamics and constituents of the cluster.
While the components of the cluster, which is the aggregate of the behavior of the
agents is expressed as the characteristic of the cluster itself, membership in a cluster puts an individual agent within a group of interacting agents, and a common environment.  The interaction of other agents with the environment modifies the environment of the agent as well, and shared resources and the requirement for collaboration and communication also makes the other agents prone to convert the agent along their own preferences.   The need for an agent to collaborate with others for one of his preferences may cause another one of his preferences to be in connection with dissimilar states of other agents causing some of his own preferences to change to match others'.

The organization and size of a cluster and its stability may have a large degree of variability compared to the degree of fuzziness possible for an agent, and clusters also may display many degrees of complexity (subclasses), starting with simple family units and then expanding to complex social structures with labor distribution and differentiation, yet accounting for its components' individual identities. The exchange of components between cluster may be subject to more complexity and plasticity because of the spread and nature of adhering forces for a cluster and the degrees of freedom possessed by the component agents.

The agents in a cluster form bonds of matching variables, and also bonds of complementary symbiosis within that larger identity.  However, the agent's own identity is reflected in his membership of the larger cluster because the agent would identify with other agents with similar variables. Hence, the agent's own affinities with other agents would in turn give rise to cluster identities that would exhibit these binding parameters on a larger scale.

\subsection{Modification of Utility Functions}

A trust-based departure from Nash equlibria (Ostrom 1997) can be explained by modifying the utility function for an individual.
Our model takes into account a broad range of similarity based interactions to construct an extended notion of {\it self} (Shafee 2004), and the agent is rational in the sense that he tries to maximize the utility of this weighted idea of self he fits himself within.

The individual utility function can get distorted because of this affinity factor, as the corrected utility would be the utility of the individual modified by a weighted utility of an affinity group. The affinity factor will reflect an agent's perception of other agents' variable states. It is interesting to note that the agents may or may not have knowledge of the proper state of these variables. Therefore, each variable in an agent has an actual value, and also a value perceived by the other agents.  So, the affinities may not be commutative (i.e. order independent), and the gain from a perceived affinity may not be balanced by the sharing an actual affinity.

How higher order terms come into play within any cluster and the total interaction is dependent on the value of the higher order terms, and may differ from one scenario to another based on the total number of agents, the partitioning of matching variables, and the weights assigned to the variables at a specific time and place.
\begin{equation}
U=U_0 + a_0 U_1 + a_1 U_2 +  a_2 U_3 + ...
\end{equation}
Here $U$ is the utility. The first term is the utility of an isolated individual agent, while the second term is the utility from the second level affiliation of the individual with one other agent, the third being the result of interactions among three, and so on.

Although in some cases the basic laws of classical economics can be retrieved in the zeroth order, when the agents are dispersed and matches are random when higher order terms can be neglected, in many other cases higher order terms are observed clearly  as is described in the end of this paper, and in instances, a higher order term may overshadow the first order term and lead to perplexing actions from an individual such as altruistic sacrifices (Shafee 2009b ).

\section{Defining Agent Variables}
Agent variables were defined in (Shafee 2004).  We broadly categorize possible variables:

\subsection{Preference Variables}
An agent's preference can be formulated as follows: The states of the environment can be imagined as a mixture of many possibilities that can collapse to only one of many possible states with which all the agents will be connected. Hence, an agent's preference would be to create a chosen environment state. Now, agents will bid for different future states that will suit their own preferences. However, an environment may be shared by agents so that only one macroscopic environment state is possible that needs to be shared by connected agents.  The acquisition and interaction of preference variables in a social network has been studied (Shafee 2009).

\subsection{Skill and Aptitude Variables}
Inherent talent or aptitude and training both come into play in the final skill of an agent. Although extreme cases prevail, some people cannot be trained in certain skills even to the level of an average, whereas some people like geniuses may excel (Johnson 2004) and produce at the top level with little or no training, for an average person a combination of ability and training is important.

\subsection{Origin of Beliefs}
	A belief consists of a set of initial states connected to a set of output states, often giving rise to rules.  A belief may be acquired by interaction with the environment or by interaction with other agents. A belief may also be imagined, in which case the belief with no corresponding environmental state can be described as a set of input states that are connected with a set of output states due to corrupt memory or mismatch within the agent. The interactions between stiff beliefs in a coherent social networks has been studied in (Shafee 2009b).

The emergence of these variables from the interplay between genetic maps and interactions with the environment is studied in detail in (Shafee 2009 and Shafee 2009b).

In (Shafee 2002), an interaction based game was proposed based on agents sharing possibly dissimilar axioms leading to their actions.  Agents interacting with one another attempted to influence other agents by converting them to be on their side. In (Shafee 2004), the notion of gambling and risk was discussed to understand how beliefs can become parts of a person's or a society's identity, although these beliefs may later become uncorrelated with a changed world scenario.

\section{Perceived Interactions and Actual Costs}

The notion of perceived identities was defined previously. We thus differentiate two types of interaction Hamiltonians.

$H_{perc}$ is the perceived Hamiltonian, which signifies the (negative) utility perceived by the agent locally, that may include virtual states as will be explained. The actual Hamiltonian $H_{act}$ excludes cues (see Raajmakers and Shiffrin 1980 for a discussion about cues) and beliefs about
external states local to an agent, that the environment cannot directly connect with.

The agent makes its moves and calculations based on his perceived interaction Hamiltonian, $H_{perc}$ which is highly localized within the agent, though a function of previous agent-agent or agent-environment interaction. The actual interaction Hamiltonian $H_{act}$  dictates the overall agent's position with respect to the current environment (i.e. the overall stability of the coupled identity components).  The terms in $H_{perc}$ are dictated by the agent's knowledge of the environment, which may not reflect the current environment state or dynamics.  The difference between the effect of an environment state in $H_{perc}$ and $H_{act}$ arises from the vastness of the environment and the many degrees of freedom within the environment.  The perception terms arise from repeated interactions with the local environment within the agent's interaction range, and involve only coarse grained interactions at the same level as the agent-environment interaction states.

Most of the time, we shall equate $H_{perc}$ with $H_{act}$, except for a few exceptional cases where an agent's actions will also depend on perceived hypothetical connections.

The dynamics of perceived identities is discussed in (Shafee 2009c) and the connections between perception and actual identities is formulated mathematically in (Shafee 2009b).

In (Shafee 2009a), more detailed mathematical formulations of variable dynamics is shown, including the truncation of the array of expressed variables because of blocking (property of mutual exclusiveness) imposed on one type of interaction by another interaction. Weights and priorities are assigned to variables as well.  Here, we concentrate on the behavior of a few important expressed variables of agents placed in groups, and see how changing various parameters update the identities and behavioral patterns of the groups and the individual agents.

\subsection{Weights and Priorities}

The weight or priority of a variable defines the degree of interaction or the fraction of total interactions expressed by an agent at a certain time.  So, high priority or highly weighted terms are expressed while low priority terms may not interact or may remain suppressed.  Again, it is the highly weighted variables that count most when relationships between agents and formation of clusters are considered.

Dynamics in the highly weighted variables, and shifts in weights are hence vital in cluster evolution.

A detailed mathematical description of weights and weight dynamics is carried out in (Shafee 2009).

\section{An Agent in a Group}
We now discuss some further examples that can be explained by using our model. How these specific situations may arise by taking self-interaction corrected economic theory into account will be briefly discussed here by pointing out phenomena that may arise because of specific interaction terms.  In this paper, we present the scenarios and points of stability graphically.  This is somewhat similar to the way supply and demand curves are used to find points of equilibrium. We will be using qualitative arguments to deduce main features for the shapes of the graphs.  These graphs represent interaction energy for different types of variables in the y axis. The origin represents a single agent who is the observer.  In the x axis, neighbors are included in order of similarity (the most similar agents are closer to the origin in order to minimize conflict and hence yield low interaction energy). The actions of the agent (eg. his formation of identities) are given by the combined interaction potentials.  Minima formed in the total interaction energy with increasing neighbors indicate formation of group identities.
Although, in this paper, we will be using qualitative arguments to define the shapes of each of the energy curves, it is possible to predict them phenomenologically by parameterizing a system in the same manner it is possible to formulate precise supply and demand curves in conventional economics.

\subsection{Collaborative and Competing Variables in a Group}

We have defined identities at the individual agent level as well as the group level.  The variables of each agent also interact within the agent, with other locally coupled variables, and with the variables of other agents and the environment.  The interaction of a variable with the environment may change (flip) an environment state to one that is more energetically favorable.  Therefore, such flipped environment states satisfy a preference of the agent.  When an agent is placed in a group in a common environment, preferences of different agents interact with the same environment.  However, the basic premise of game theory (Nash 1950) and equilibrium of supply and demand in macroeconomics acknowledge and show how multiple agents competing for a common resource compete against one another to reach various equilibria. Hence, some preferences may be such that the introduction of multiple agents may foster competition. Interaction among agents with such variables are mediated by a common environment, and result of interaction would depend on the scarcity imposed by the second agent. This competition term is similar to the result of multi-agent behavior found in conventional economics.  The other terms involving collaborative interactions and interactions in terms of genetic similarity or similarity in belief show up as corrections to the expected rational behavior.  Hence, the correction imposed by our interaction based similarity model appear as superposed peaks that take into account utility (in the form of low interaction energy) from interaction among matching variables among agents.

\begin{figure}[ht!]
\begin{center}
\includegraphics[width=14cm]{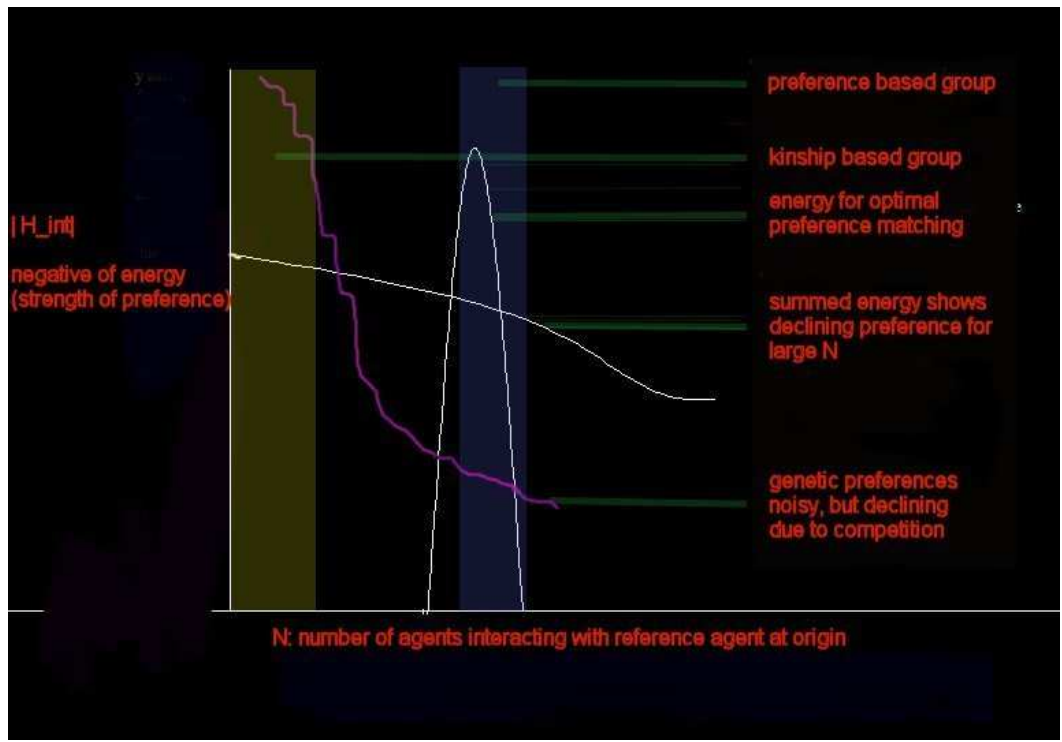}
\end{center}
\caption{\label{fig2}The effect of various types of variables within the agent's interaction potential
 }
\end{figure}

\begin{figure}[ht!]
\begin{center}
\includegraphics[width=16cm]{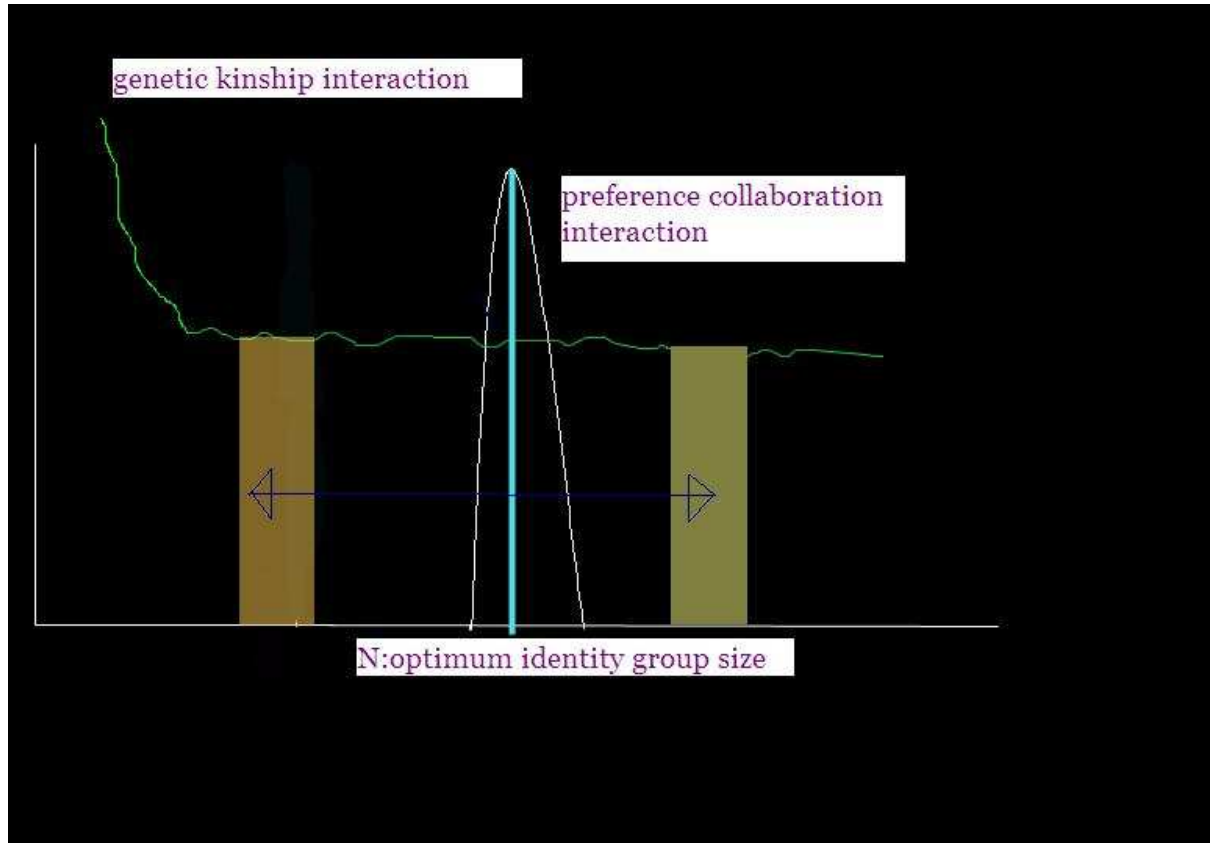}
\end{center}
\caption{\label{fig2}The problem of choosing an identity group within a large homogeneous population when collaboration requires a subset of the population}
\end{figure}

\subsection{Defining Types of Interactions in Clusters}

Since each agent is a complex entity with many variables in possibly different states, in a large group of agents, similar and dissimilar variables exist together within an interconnected network.  The vast number of variables make it impossible to have two agents that are completely similar or dissimilar, or create an absolute indicator for closeness between two agents.  Similarity in one variable between two agents does not make them absolute friends, and some similar variable states can be found even among people who can be attributed as enemies.  We thus have a weaving of similarity bonds among agents giving rise to the non-transitivity of friendship. The weight of a variable and the dynamics of the weight of a variable determine who are closer at a certain time.

The measure of closeness or the extension of the concept of {\it self} is not absolute because of the existence of competing similar and dissimilar variables and the constraints of resources.  As the number of agents increase within an ecosystem, needs for collaboration vary, and diversity in states increase. The constraints in resources and energy also put a limit to friends that can be accommodated.  Hence, societies may split and friends may become enemies.

We discuss how the total number of agents, $N$, determines the macroscopic behavior of a cluster as  $N$ is increased slowly, starting with a group of a few agents that are matching in a few variables and mismatching in some others. We also observe the effect of weighted match in the macroscopic behavior of the cluster and the creation of sub-clusters.

We take into account the leading highly weighted variables of each agent to study his formation of bonds and identities.  We also observe how the mix of competing and collaborating variables, together with variables that show a continuous degree of match contribute to interesting dynamics.

Let us take agents $S_i$ such that
\begin{equation}
{\bf S_i}=[s_i^1, s_i^2, s_i^3 .  .  .  .  .]
\end{equation}

Only highly weighted components of the self-array of agent {\it i} are expressed in the interactions, and we observe how these expressed terms affect the macro interactions.

To start with a simple picture, let us assume that there are only {\it m} highly weighted variables.
Of these, we define $s_i^1$ as a survival variable, so that the agents must compete to connect these with the environment.  For example, it can be food, so that connections to food has a very high weight. However, if more agents are introduced in the picture, each agent has less food and the term $s_i^1 . h^1$, which is the connection of an agent with a food, can occur only less frequently since other agents need to connect with all available food states.  Such food states, after interaction, are converted into unusable smaller states, so that the total number of available food states would determine the availability of food to an agent.  The high weight of $s_a^1$, is fixed, and would require the agent to find a food state with a high frequency. So, the introduction of more agents would it uncomfortable for agent {\it i} with respect to competition based survival terms such as $s_i^1$.

The next  three variables are markers and preferences that the agents may be in matching or mismatching positions from agent to agent forming matching bonds. Let us assume that $s_i^2$ is genetically fixed, and the other two are in state 0 or 1 with a certain flipping cost.

Let $S_i^6$ be a skill based collaboration variable. For simplicity, we now assume that all agents have the same skill level. However, a collaboration with $N'$ agents is required for the production of $M_{P_6}$, the optimal environment. Hence,  $S_n^6.h$ (the interaction of $S_n^6$ with the matching component of $h$, the environment) will  have a peak when $N'$ agents are introduced in the neighborhood. Such a peak may result from an optimum value of collaborators for the production of the related skill-based item. This peak can be a Gaussian with the top at $N_{skill}$ that also has a cut-off so that it falls to zero if $N_{skill}<N_{thresh}$.  This signifies that less than $N_{thresh}$ agents collaborating will yield no macro output.

\subsection{Expression and Suppression of Variables in Cluster Dynamics}

If $N$, the number of agents, is small, fewer agents compete for a fixed amount of food. In this case, it is possible to have the contribution of the competitive term, $s^1$ , decrease slowly in the interaction landscape as $N$ increases. The term, $s_N^m$ , may require $N'$ agents with mismatching  variables to collaborate, so that the state  $M_{P_m}$ can be created. If the yield of $M_{P_m}$ interacting with a certain variable of an agent is large compared with the effects of the other mismatching terms lumped together, a peak will occur near $N'$ and  $N'$ agents, even though mismatching in attributes, will collaborate and find a common identity based on mutual need in  $S_N^m$. Hence, it is possible to observe friendship between diverse agents when placed in an environment where collaboration effect is large and the total number of agents available for collaboration is small.  However, as the number of agents is increased with a few expressed parameters within the cluster, splits within the cluster can be observed.
An example can be diverse and even conflicting ethnic groups forming bonds to collaborate against a common enemy by forming an alliance when coherent mutiny or non-cooperation is needed to fight against a common enemy.  If the weight of the collaboration term is high, the ethnic tags and cultural differences (which are the middle three terms in the array just described) will gain low weight and can remain suppressed.  An example can be the formation of a common black identity in many colonial countries where many conflicting tribes formed alliance.  However, when collaboration is not needed or the common enemy is removed, so that the last parameter requiring collaboration is taken away, the first parameter, where competition among individuals for food is expressed gains prominence and the collaboration can fall apart, creating subclusters based on the previously unimportant parameters, which can be genetic tags or certain cultural distinctions.  A case study of the breakdown of black identity in South Africa was discussed (Shafee and Steingo 2008).

Another interesting example would be several ethnic groups being grouped together by socialism, so that a coherent culture is superficially created suppressing individualism.  The inertia acting for and against flipping to individual ethnic identities and keeping coherence can be modeled mathematically as follows:

The suppression of individual identity can be brought about by the need to collaborate to cater to the basic preferences.  It can be furthered by the achievements leading to security produced by collective group production.  However, the inertia against a state of mass unification is posed by genetic taste, genetic identification among sub-clusters and kinship, and mismatch between preferred job accounting for personal preferences and ability together with opportunity and optimal job for the society based on ability and optimal efficient need.
Historically, the rise of socialism has been observed when the fragmentation of a society leads to breakdown of economies, or when a large unskilled population that is scattered is needed to come together and collaborate to a significant degree to forcibly introduce collaboration and investment of skills for the greater good of a group (see Marx and Engels 1998 for a discussion of the communist manifesto to bring together un-united workers for needed large scale collaboration in a  fragmented and frustrated society) and frustration about socialism and the breakdown of socialism causing even genocides and ethnic cleansing have been observed when socialism is efficient in the sense that the total production of the society exceeds the bare minima and creates security for the entire society against other societies (see Remnick 1994 for the collapse of communism due to frustration about individual identities).

\subsection{Increase in Variable Numbers in Cluster Dynamics}
Let us include more preference expressed variables in each agent so that
\begin{equation}
{\bf S} ~[s^1, s^2,s^3, s^4, s^5,.... s^m,  \sum s^b:(b >m)]
\end{equation}

The term $\sum s^b$ is a background component formed by adding up all the remaining newly introduced variables in the agent.  We assume that the weights of these terms are lower than the weights of the first five terms so that all these new terms can form a background.
If $N$, the number of agents, is increased, given the condition that each agent  differs from the others at a few random places in $\sum s^b$,  an undefined arrangement of subgroups can be seen for the following reasons:

The existence of competition due to $s^1$ and the need for collaboration because of $S^m$ would tend to define the size of sub-clusters that can act as surviving units. However, bonds between matching preferences and repulsion between mismatching preferences would tend to be defining factors about which agents should be grouped together. $N'$ (the number of interacting agents) and the number of mismatching random preferences can be chosen such that an agent is on an average the same degree of mismatching from any other agent, but the mismatching would arise in different preference sites. Since most preferences can be switched at some cost, the fixed matching state would be $s^2$, which is genetically fixed, and the degree of match or closeness in $s^2$ would produce semi-stable clusters. These would fluctuate because of the changing effects of the other preferences and the small size of a sub-cluster where the second term has a large degree of match.  If the number of agents required for collaboration is larger than the sub-cluster formed by matching the second term, but still much smaller than the total number of agents found in the entire cluster, the total number of agents required for collaboration can be unstable or unviable.  Attempts to increase membership based on the genetic kinship term may increase the total number of agents within the cluster as each kinship subunit extends its membership.  This might lead to scarcity, making the repulsive competition term much steeper, causing large degrees of competition for resources within the cluster.  The attempts to include random members of the cluster to build these preference sub-clusters might be unstable because of the lack of diversity within the cluster. As a result, the average match of any two agents disregarding the genetic match is about the same, and any member can be replaced with another member from the large cluster. This situation can lead to the instability of the preference sub-clusters.

Now let us assume that instead of random slight mismatch, we have specific variables that can be mismatched, while keeping the rest of the variables steady.
So each agent has a property  array
\begin{equation}
{\bf S_i}=[s_i^1,s_i^2, s_i^3, s_i^4, s_i^5, ... s_i^m \sum s^b]
\end{equation}
And in the total cluster, $N'$ agents have, due to the newly introduced background, a pool of aggregate variable states
\begin{equation}
\sum_{i= 1..N'}s^b_i
\end{equation}

These newly added variable components change only to contribute a background noise and the effects of more prominent terms like $s^1$ can be studied by superposition of the interaction due to $s_i^1$ onto a background.

\subsection{Subgroup Competition in Scarcity}
As the population, $N$, goes up, it is possible to form larger group identity subclusters within the initial group, each subgroup attaining an identity due to  marked difference in a  highly weighted variable. Let us say, for example, that $s^3$ has a high weight after $s^1$ and $s^m$.

It is possible to divide $N$ into subgroups depending on $s_{i_M}^3$ if it exists in two distinct separate states because of historical mixing of two populations or mutations or separate histories created because of long-lasting ideological differences. If they are all in the same environment, the once friendly agents may now form their own competing subgroups, who are rivals (Flint, 2006).
Let us denote two subgroups, $A$ and $B$, comprising agents such that each agent also comprises  an array of variables within the cluster:
\begin{eqnarray}
A= [A_1, A_2 , ....]  \\
A_i =[s_{A_i}^1, s_{A_i}^2,0, s_{A_i}^4, s_{A_i}^5,s_{A_i}^m, \sum_bs^b_{A_i}]  etc.
\end{eqnarray}

Each of these subgroups will attain an identity that is dependent on the value of $s_{X_i}^3$ ($X=A,B$), given the other values produce random noise with a low weight.

When still contained within the same cluster because of the availability of resources and opportunities, this type of splitting will create two different identities within the same cluster that may compete for the common resources.  The coexistence of two identities that need to share the same resources would create competition within the cluster.  Let us assume that the total population of the entire cluster is slightly less optimal for the total resources available, so that the total resources cannot cater to all the agents properly. In that case, the survival competition term will be steep if all agents are combined, but may not be steep if only one subgroup exists, while the collaboration term would require a number between one and two subgroups.  If the skills required for proper utilization of the resources are spread within the two sub-clusters so that no sub-cluster can utilize the resource properly, the polarization based on identities may cause inefficiency in utilizing the resources. Each of the subgroups tries to exclude the other to keep the survival competing term less steep, but may need certain agents from the other subgroup to properly utilize the resources.  This situation may create clan strife within the cluster, or the need to find a common identity to suppress the split of the cluster.  Forcible nationalism or socialism to suppress divisions among such clusters consisting of several subclusters may again lead to frustration in individuals if superficial common identities (Remnick 1994).

These subgroups will then form a unit that will compete for the resources for their own conflicting preferences. As competition gets tough, $N_{skill}$ may go up as more skill is needed to obtain the same resource under competition.  So, each group may alternately try to increase members within itself in order to become self-sufficient in skill, and then exclude the other subgroup from the common resource. Examples can be given of rivalry between almost similar ethnic groups in highly populated areas with scarce resources (Ahmed 1991; Bhabani and Becker 1999).

\begin{figure}[ht!]
\begin{center}
\includegraphics[width=14cm]{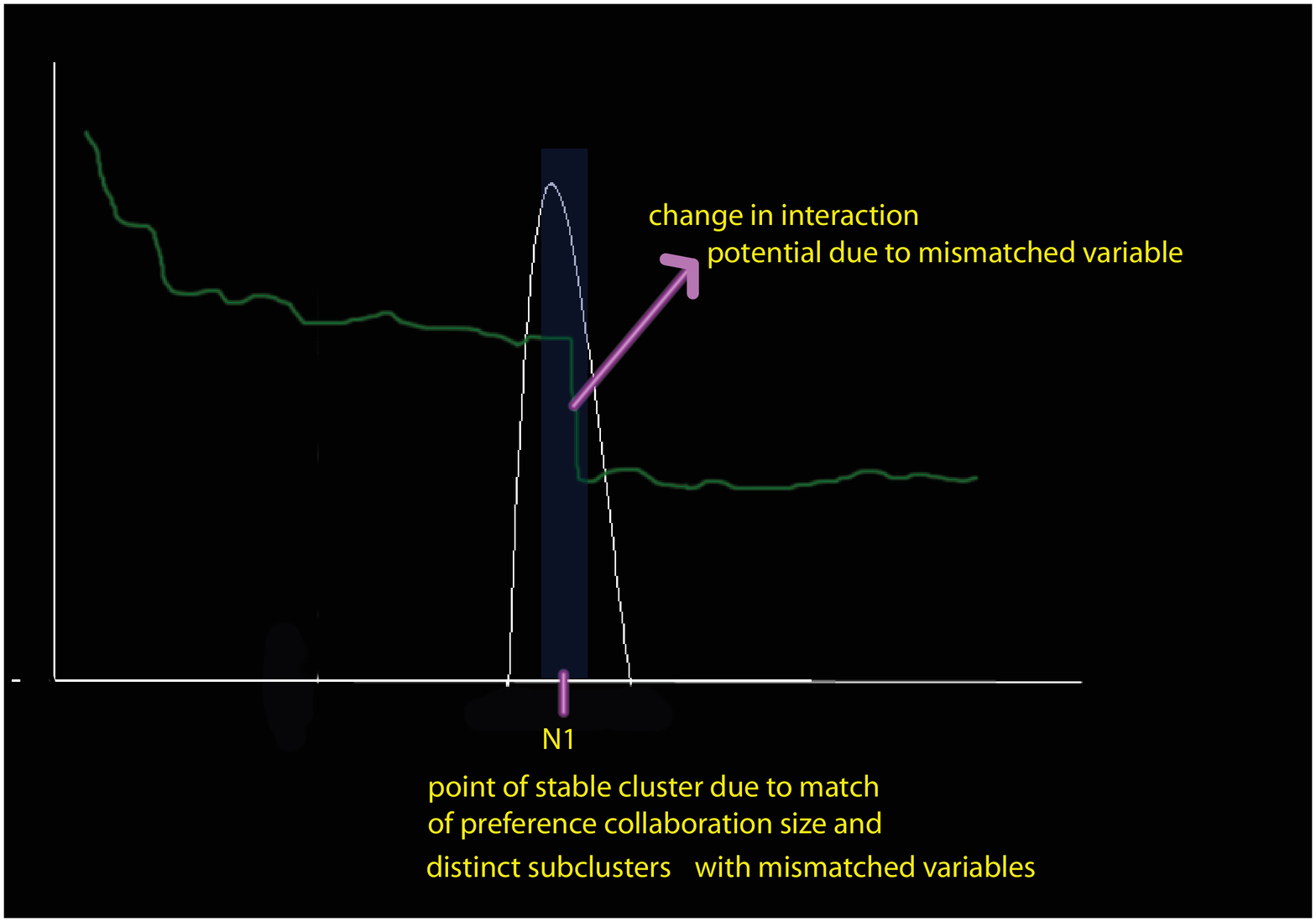}
\end{center}
\caption{\label{fig3}Almost stable group identities deriving from coinciding collaboration utility and matching subcluster size}
\end{figure}

\subsection{Stability of Subgroups in a  Large Population}
	Another limit point would be a large population where many preference collaboration groups are possible.  However, the match-based energy landscape has no single distinct marker or step; so these groups can have members exchanged without incurring a high cost.

\begin{figure}[ht!]
\begin{center}
\includegraphics[width=14cm]{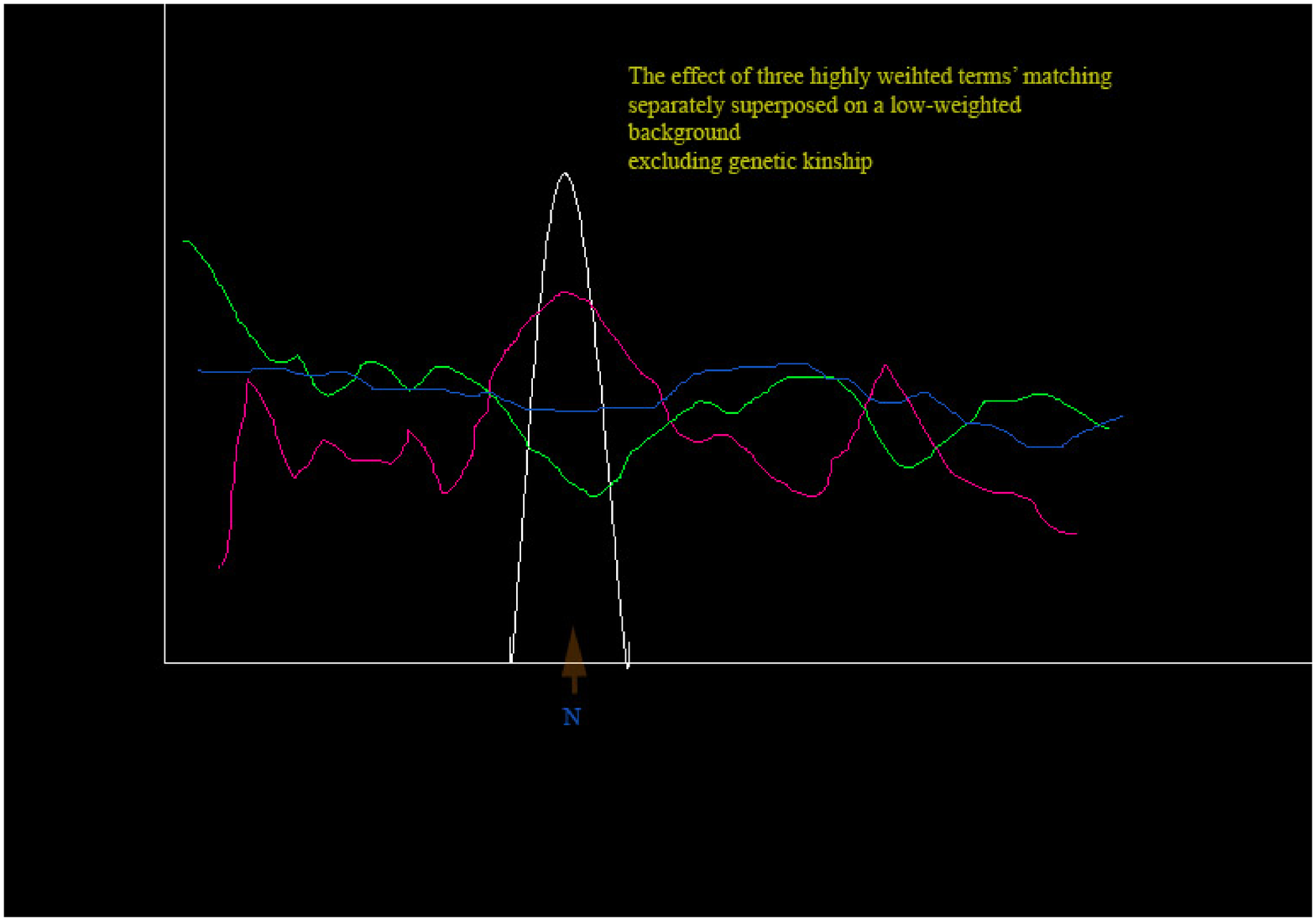}
\end{center}
\caption{\label{fig4} Three different match interaction terms summing up to almost constant background potential.  Genetic kinship has been neglected.}
\end{figure}

As a result, agent ${\bf i}$'s possible identity group, which has agent ${\bf j}$ in it, might be different from agent ${\bf j}$'s possible identity group. It is then possible for each agent to have more than one preference identity group, with agents interchanged, if two possible identity groups (Each containing $N$ agents needed for preference $P$ are almost homogenous, but only $N$ agents are needed for preference $P$.) In that case including the $N+m$ th agent will have a diminishing effect, but the $N^{th}$ agent can be chosen from more than one candidate.  In these circumstances, highly volatile groups will form, and stable matching, such as genetic kinship will have a longer stability because although other preferences can change, these stay fixed. With genetic kinship fixed, and the need to have $N$ agents for preference $P$-based identity, the agent will have an urge for a larger number of genetic kinship-based collaborators.  So, societies of this type, where there is a large degree of random mismatching in many variables, with no certain distinct markers, large families will be formed and there will be a tendency to retain family members in collaboration projects. If there are less than $N$ kinship members available, the preference  based identity clusters will be volatile. This situation can be somewhat similar to large degrees of ethnic nepotism in slightly diverse high population areas (Vanhanen 1999, Ahmed 1991)

The competition to increase the share of each group, and to enhance the size of the group in order to have a larger fractional share, leads to a higher total population and a more acute severity of the scarcity in a manner matching the "Tragedy of the Commons" (Garrett 1968).

\section{A Specific Case Study}

\subsection{Analysis of Characteristics of the Russian and the French Revolution}
In this section, we analyze and compare two major episodes of history to qualitatively see how using a variable-state interaction based social model is appropriate in fitting many social phenomena.  We especially note the emergence and evolution of identity classes and their interactions, as well as stages of social stability and new social identities reached as a result of these interactions.  We especially note how collaboration clusters are formed, and how numbers and similarity factors become important in the formation of such energetically favorable identity units.  Facts used in this section can be checked at (Sakwa 1999; Crankshaw 1978; Ascher 1994; Medvedev 1979; Fischer 2001; Perrie 1976; Redkey 1958; Alexandrov 1966; Carlyle 2002; Furet 1995)

The two specific examples we will use here are the formation of the USSR and the French revolution.  Although both had some similar features, leading to the collapse of {\it ancien regime} the outcome in one led to the communist regime, while in another, a republic was created that also saw periods of different types of political identity later.

The {\it ancien regime} was characterized by marked stiff social hierarchy with the monarch holding absolute power over enacted laws. Hence, identity classes in the society were separated, although as not markedly segregated with ethnic tags as in the case of South African apartheid. However, in the case of Russia, ethnic divisions and conflicts also took place in an inhomogeneous society.

The identity segregation into sub-clusters within the societies in {\it ancien regime} was based on position that usually passed down the bloodline of families and not large ethnic subclasses. Monarchs were challenged by contesters, and wars broke out between contending bloodlines, keeping much of the classes still segregated.

The identity of an individual hence involved an identity which was part of an entire system with a specific overall history, and also a subclass of the system based on its social position.  A clergy class was officially placed to deify the monarch as part of an integrated hierarchical social identity. The army protected the monarch, making the act of questioning the authority a high risk task, separating the identity clusters.  Each identity class, thus had preassigned economic identities as well, separated mostly by genetic lineage.  Individual interactions were concentrated within identity classes, which interacted among themselves as large hierarchical subclasses with different identities.

Collaboration within each of these subclasses somewhat modified the cultures within the subgroups, isolating one another further. In our previous graphs of collaboration-based identities somewhat similar people formed strong identity groups to collaborate. In this case, similarity within stiff classes was often imposed by holding members fixed for generations to pass down a culture connected with the avocation.  Hence, subclasses were initially formed and then genetic kinship was correlated to avocations to create cultural subclasses that were held together by similarity and continued to collaborate in similar avocations.  The success of such such identity groups depended on the possibility of training an average member of any cluster to perform the task related to that cluster with at least considerable success.  Such mechanism also resulted in the formation of traditions that were seen parts of identities.

Collaboration cultures based on economic activities also modified total output products and available goods within each subclass, further isolating the subgroups.  Hence, the rational behavior of an agent was somewhat dictated by his identity class because of the shifted supply and demand possibilities within each class. Again, it was genetic affinity and not always skill that helped maintain the classes that required a steady membership of large numbers of people affixed to certain cultural subgroups.  Not all monarchs were great rulers, and not all peasants were low risk takers. However, membership within an identity class held together by genetic lineage suppressed easy mobility due to skill. Each cultural subgroup hence had preferred skills based on the identity of the subgroup that got expressed.

The monarchs often chose their consorts from different countries and cultures because of direct foreign interactions and need for forming collaborations, making the upper and upper-middle class culturally separate from the peasants.  For example, in Russia, the upper and upper middle class spoke French and German as opposed to the peasant class speaking Russian.  When the Narodniks, who originated in mostly {\it enlightened} upper middle and middle class, and sought to recruit peasants for a revolution, went to the villages, cultural differences caused the peasants to even consider them as witches.

In both France and Russia, the revolutions and large scale social changes started with revolts within social groups after economic woes due to war and famine.  In both societies as well, the revolution originated not within the isolated peasant or working class, but either within the army, which contained both titled officers and peasants recruited as soldiers working side by side, or dissatisfied upper or upper middle class.  In France, a foreign financier named Necker was a key instigator of the revolution, who proposed that taxes be imposed on the nobility as well, and in Russia, the Decembre Revolt was initiated by the dissatisfied army, and the liberalization of an upper-middle class and middle class who were dissatisfied with the monarch and prevailing conditions.  Again, the Russian revolution was fostered by the social evolution leading to the creation of a industry based working class that had greater access to city life, and the uprise of a professional middle class whose roots could be founded within the peasants.  The dynamics created by changed economic needs of the society in competition with other countries caused mixing and overlap of identities fostering interaction among segregated identity groups.  Lenin, who was the leader of the Bolshevik group, that believed that the revolution must be achieved by the industrial working class was a practising lawyer with access to modern education and international connection who was born to a school inspector and had mixed ethnicity originating in several countries.

In both France and Russia, the revolution took place in steps, with a number of groups representing power to different social classes and identities challenging the monarchy.  In both countries, the revolution saw periods of unstable equilibrium solutions that were challenged by factions representing separate identity groups.

In Russia, the emancipation of serfs was followed by revolts from the Zemstvo (liberals, asking for constitutional monarchy), the Socialist Revolutionary Party (opting for land distribution among peasants and an agrarian peasant based revolution; included membership by many upper middle class and middle class revolutionaries) and the Marxist Russian Social Democratic Labour Party (opting for nationalization and Marxism with the idea of industry workers as the revolutionary class). The first semistable solution was the creation of the State Duma, satisfying the liberals.  However, infiltration within the SRP by a police spy led to its split into two parts, one joining the SRP and the other supporting the second semi-stable stage -- the creation of a provisional government in collaboration with the Duma and the socialists. The SDLP itself was
divided into Mensheviks and Bolsheviks based on their disagreement on membership policy and whether Russian industrial workers were prepared to stage a revolution or if such revolutions should be preceded by a bourgeois revolution.

In France, similar factions formed after the initial moves were made for liberalization.  Besides, Necker, Emmanuel-Joseph Sieyès, who was a key provoker, held a position in the a clergy, and hence was a member of the privileged class, who had some noble blood, but was brought up a commoner.  Although he derided the old fashioned school of ideas, and was expelled from school once he received his degree in theology and received a position within the clergy in spite of his contradictory views by using powerful friends.  This depicts the complexity of the number of variables in interactions among individuals when optimal bonds formed based on certain variables can give rise to overall friendship that can be used to influence to procure a position in an opinion group regardless of contradictory opinions in another variable.  However, this position was later used in bargaining for rights that went against the clergy.  Pamphlets like {\it Qu'est-ce que le tiers état?} eventually led to the clergy losing their privileged position.

Other instigators of the French revolution involved wealthy professionals.  The already existing strong bourgeois class in France was aware of international trends.  This was unlike in the case of Russia, where 83 percent of the population comprised the peasant class and exchange of ideas at the international level was concentrated within a small fraction that was able to move to the professional class. After the revolution, the workers and the peasant were isolated from international contact by nationalization, information was filtered and private ownership or entrepreneurship was banned. In France, the peasant class did not need to be part of a revolution to topple the monarchy, and the peasant class was involved only after the revolutionaries took over.  They fought against the revolutionaries because of their favorable relationship with the clergy, since religion was a part of their identity, and was later crushed brutally by the revolutionaries.  Moreover, the revolutionaries banned workers' organizations.

France also saw several stages of semi-stable equilibria, starting with the creation of a unified assembly, the fear of being persecuted by the army leading to the assembly's attempt in creating a new constitution and the fall of Bastille, attempted defection of the Royal family, constitutional monarchy, then the failure of the constitutional monarchy making way to the ultimate execution of the King and the Queen and for a reign of terror involving factions, and then the creation of a directory.  Such episodes of unstable identities in both Russian and French histories involving the succession of power from one identity group to another point to the processes of removal of unfavorable interactions and sustenance against unfavorable states within a larger system with subgroups of identities existing in a hierarchical manner.  Strengths of each identity class, bargaining power derived from skill correlated with them and mobility between identity classes are also important and will be analyzed with respect to our model in the next section.

Three specific points associated with the events in these revolutions are worth analyzing before we move to the next section where we qualitatively formulate our theoretical model.

The split of SR into supporters of the duma and supporters of Bolsheviks show the importance of group identities in a large scale social movement and how an ensemble of individuals with wide spectrum of views can be aggregated within one identity class, where they are held together with group values they adhere to but may not completely agree with.  However, when one such group dissolves, each member makes a choice to join the next possible best match of group identity, which might push members from two ends of the opinion spectrum within the same identity group into two conflicting groups, making once friends bitter enemies.

Again, each country, consisting of social classes is placed in competition with the rest of the world, especially other neighboring countries.  When the Russian revolution started, the French revolution had already taken place and Britain was a crown monarchy.  Many European nations were more industrialized than Russia.  However, Russia's position within Europe made it necessary to start shifting towards industrial scenarios, thus creating a working class that came in contact with city life and to some extent the international scenario.  However, after the revolution, complete nationalization and overthrow of the entrepreneurs made it possible to isolate the large population that was unaware of world trends isolated again.  Competition with he world was achieved by having a select few people choose homogeneous policies for the entire country that was competitive with respect to world progresses, making USSR pioneers in many fields of science. Initial increase in living standards, compared to the commodities and security available to the former working class and peasants helped overthrow the non-Bolsheviks and created a large scale unified economy. However, de-personification of the individual gave rise to high mortality rates, lack of goods of individual choices within the country, and dissatisfaction, eventually leading to Perestroika.

However, after Perestroika, a sudden introduction of the free market, with not much of the infrastructure changed, posed the necessity of making choices and decisions at a personal level in a competitive international market. To an economy that had been closed for a long time, this resulted in considerable chaos.

The last point to consider is the incompleteness of information regarding each individual in an ensemble, where the individual himself is a complex array of variables, leading to the uncertainty in strength and motives of another identity class, and inertia associated with different levels of identities, especially when the dynamics is fast.  Initially, Nicholas was to be tried by the Bolsheviks, but his entire family was murdered when the Tsar's supporting white army was heard to be approaching to rescue him.  In France, King Luis XVI was executed for treason, even after the country attempted constitutional monarchy, when Austria, where Marie Antoinette came from, declared war, threatening to restore the King.  In France, the Tennis Court Oath took place because the assembly members were unsure if the army, consisting of mostly hired foreigners, was approaching to persecute them, giving rise to instability.

\subsection{Theoretical Model of Interacting Hierarchical Identities}

We use the analysis from the previous sections to propose a theoretical interaction based hierarchical identity model to be used later in conjunction with detailed mathematical formulation:

1.  Each agent consists of an array of a large number of variables in different states.  Each agent has also preference for environment states and material goods.
2. Collaboration and need of skills make it necessary for agents to cooperate and share.  Hence all agents cannot have all needs satisfied.\\

3. Collaboration for different tasks is often based on and sometimes evokes notions of similarity, and gives rise to identity classes within a society.  These classes may have various degrees of dynamics or overlap.\\

4. The top class of the hierarchy level, and the top person at a hierarchy level has more of preferences satisfied.\\

5. Each person tries to reach the top.  However, this is subject to his bargaining power, ie. skills within his array of variables.  If he has a specific skill correlated with his individual qualities that is not found in other agents, or other groups, he as a whole has a larger bargaining power, and can place himself higher.\\

6. Within a society, collaboration is needed for specific tasks making it necessary or create identity groups. Groups need to be formed to challenge an existing collaboration group to interact at a group level.  Groups have group bargaining power dependent on the skill and number of people within the group.\\

7. Group identity can depend on similarity of variables, sometimes genetic, sometimes beliefs or other types.  However, one agent may move to a different group because of match of specific interest or skill, although that agent may still identify with his group of origin.\\

8. Mobility of agents among groups, hence scale-wise hierarchical interaction causes information leakage between classes.  Often the position of an agent in one group can be used to mobilize a different group the agent identifies with.  The mobility of an agent to a separate group is often by means of the skill variables needed within a group to maintain its bargaining power, and also due to interactions in many weighted variables while local needs may make one variable weighted (more important) at a time, though priorities may later shift.  Shifting of such priorities is subject to plasticity of the variable.\\

9. Shifting of power between groups may take place in stages with unstable equilibrium points.  Shifts depend on breaking unfavorable interactions and also being able to sustain against other groups from a new position.  Often the shift starts with compromised joint positions between the top group and a group closer to it, but as unrest persists, mobilized by shifts in identities in other groups as well, dynamics persists, until favorable interaction scenarios subject to balance between bargaining power and optimal interactions are reached.\\

1o. Some of the identity groups are stiff and some others are more flexible.  However, some degree of inertia and stages of dual identities can be observed in most dynamical shifting.  Also, identities can split, subject to changed interactions.  Trust and risk in isolating an agent's identity or the number of agent's in an identity group may lead to extreme behavior.

In (Shafee 2009a) a quantitative formulation of some of these features is shown based on hierarchical interactions among identity units in a complex organization.  The concept of a system consisting of interacting subsystems dynamically approaching minima in an energy landscape is elaborated more rigorously in that paper.

\section{Conclusions}

We have discussed the various forms of identities that can originate due to interacting variables in semi-closed systems on different scales.  The introductions of different levels of identity lead to new forms of economics more consistent with real social behaviors which may often appear irrational. We have outlined the possibility of developing concepts and relations related to the evolution of social clusters, which are  analogous to spin systems, and have indicated the importance of the concept of the identity of each agent. Variable attributes, which may be quantifiable, are specified to define levels of identity.
We have then used the concepts developed earlier in the paper to study the dynamics of social clusters in specific cases and have graphically found some critical points in the evolution of social clusters. More mathematical applications of the concept may be able to predict of real social situations.

\section*{Acknowledgement}
The author is thankful to Kevin Mitchell for technical help.
\section*{Bibliography}

Ahl V and Allen TFH, Hierarchy Theory. New York: Columbia University Press; 1996.

Ahmed S, The Politics of Ethnicity in India. Regional Studies (Islamabad). 1991; 9(4): 22-5.

Alexandrov V, 'The Sokolov Report'  The End of The Romanovs. London: ;Hutchinson; 1966.

Ascher A, The Revolution of 1905, vol. 2: Authority Restored. Stanford University Press, Stanford; 1994.

Barabási A and Reka A, Emergence of scaling in random networks. Science. 1998;286: 509-512.

Bhavanani R and Backer D, Localized Ethnic Conflict and Genocide: Accounting for Differences in Rwanda and
Burundi. Santa Fe Institute Paper 99-07-053.1999. www.santafe.edu. 1999.

Black F,  Scholes M, The Pricing of Options and Corporate Liabilities. J. Pol. Econ. 1973; 8 (13): 637-654.

Blake R, Ramsey G, Perception: An Approach to Personality. Ronald Press: New York; 1951.

Bowles S and Gintis H, The Evolution of Strong Reciprocity: Cooperation in Heterogeneous Populations. Theor. Pop. Biology. 2004; 65(1): 17-28.

Bowles S  and Gintis H, Persistent Parochialism: Trust and Exclusion in Ethnic Networks. J.  Econ. Behav. and Organiz. 2004b; 55: 1-23.

Bowles S,  Individual Interactions, Group Conflicts, and the Evolution of Preferences. Santa Fe Paper 00-08-047. 2000;  www.santafe.edu. 2000.

Bunyan J and Fisher HH  eds, The Bolshevik Revolution, 1917–1918: Documents and Materials. Stanford: Stanford U. P., 1961; first ed. 1934; 1961.

Carlyle T, The French Revolution: A History. New York: The Modern Library; 2002.

Connolly K  and Martlew M, 'Altruism', Psychologically Speaking: A Book of Quotations. BPS Books; 1999: 10.

Crankshaw E, The Shadow of the Winter Palace. London: Penguin; 1978.

Dawkins R,  The Selfish Gene. Oxford: Oxford Univ. Press; 1989.

Doyle, W,  Origins of the French Revolution. 3rd ed., Oxford: Oxford University Press; 1999.

Edwards S and Anderson P, Theory of spin glasses. J. Phys. F. 1975; 5: 965-974.

Epstein JM , Learning to be thoughtless: Social norms and individual computation. Computational Economics. 2001; Vol 18(1): 19-24(16).

Esch T and Stefano G, The neurobiology of pleasure, reward processes, addiction and their health implications. Neuroendocrinology Letters. 2004; 4:25.

Fallon JH, Keator D, Mbogori J and Potkin  SG, Hostility differentiates the brain metabolic effects of nicotine.  Cognitive Brain Research. 2004; 18: 142 -148.

Fischer L,  The Life of Lenin. London: Orion Publishing Co; 2001.

Flint  J and  de Waal A,  Darfur: A Short History of a Long War. London: ZedBooks; 2006: 25

Fudenberg D and Tirole J, Game Theory. Cambridge: MIT Press; 1993.

Furet F, Revolutionary France, 1770-1880. Malden, MA: Blackwell Publishing; 1995.

Gastner MT and Newman ME, The Spatial Structure of Networks. European Physical Journal B. 2006; 49(2): 247-252.

Gell-Mann M, The quark and the jaguar: adventures in the simple and the complex. W. H. Freeman: New York; 1995.

Gibbs WW, The unseen genome: gems among the junk. Scientific American. 2003; 289(5): 46-53.

Gintis H, The Hitchhiker's Guide to Altruism: Genes, Culture, and the Internalization of Norms. J. Theor. Biology. 2003a; 220: 407-418.

Gintis H, Solving The Puzzle of Prosociality. Rationality and Society. 2003b; 15:2.

G\"{o}del BM,  On formally undecidable propositions of Principia Mathematica and Related Systems. London: Dover Pub; 1992.

Hamilton WD, The genetical evolution of social behaviour I and II. J.  Theor. Biology. 1964; 7: 1-16 and 17-52.

Hardin G, The Tragedy of the Commons. Science. 2007; 162 (3859): 1243-1248.

Häsler, J Samuelsson, T and  Strub K, Useful 'junk': Alu RNAs in the human transcriptome Cellular and Molecular Life Sciences. 2007; 64: 1793-1800.

Henrich J,   Boyd R ,  Bowles S,  Camerer C,   Fehr E,   Gintis H  and McElreath R 2001a, Cooperation, Reciprocity and Punishment in Fifteen Small-scale Societies.  American Economic Review. 2001a; 91: 73-78.

Henrich  J,  Boyd R, Bowles S,  Camerer C,  Fehr E, Gintis H,   McElreath R,  Alvard M,  Barr A,  Ensminger J,  Hill K,  Gil-White F,  Gurven M, Marlowe F,  Patton  JQ, Smith N and Tracer D 2001b,  Economic Man in Cross-Cultural Perspective: Behavioral Experiments in Fifteen Small-Scale Societies.  Santa Fe Paper 01-11-063. 2001b.

Hume D, Enquiries concerning Human Understanding and concerning the Principles of Morals. Reprinted from 1777 edition, Third Edition, L. A. Selby-Bigge (ed.). Oxford: Clarendon Press. Sect. XII, Part III; 1777: 165.

Imarisio S, Carmichael J, Korolchuk V, et al, Huntington's disease: from pathology and genetics to potential therapies. Biochem. J. 2008;  412(2): 191–209.

International Human Genome Sequencing Consortium, Initial sequencing and analysis of the human genome, Nature. 2001; 409: 860-921.

Iyer S, Borooah V and Dubey A, The Effectiveness of Jobs Reservation: Caste, Religion and Economic Status in India. Development and Change. 2007; 38: 423-445.

Johnsen SK,  Identifying Gifted Students: A Practical Guide.  Waco, Texas: Prufrock Press, Inc.; 2004.

Klein SB, Cosmides L,  Murray ER and Tooby J, On the acquisition of knowledge about personality traits. Social Cognition. 2004;  22(4):67-390.

Lazarsfeld PF and Merton RK, Friendship as a social process: A substantive and methodological Analysis, in  Freedom and Control. In Modern Society. Ed. M. Berger et al. Van No strand; 1954: 18.

Lewin, Roger 1992, Complexity: Life at the Edge of Chaos. New York: Macmillan Publishing Co; 1992.

Lovallo D, and Kahneman D, Delusions of success - How optimism undermines executives' decisions. Harvard Business Review. 2003; 81(7): 56.

Makalowski W, `Genomics: Not Junk After All. Science. 2003; 300(5623): 1246-1247.

Marx K and Engels, The Communist Manifesto, introduction by Martin Malia (New York: Penguin group); 1998: 35.

Mendelsohn O and Viczianry M,  The Untouchables, Subordination, Poverty and the State of Modern India. Cambridge University Press; 1998.

Medvedev R, The October Revolution. 1979; New York: Columbia University Press.

Moody J and White DR, Social Cohesion and Embeddedness: A Hierarchical Conception of Social Groups. Santa Fe Working Paper. 00-08-049; 2000.

Moreno Y,  Nekovee M and  Pacheco AF, Dynamics of rumor spreading in complex networks. Phys. Rev. E. 2004; 69: 066130.

Nash J, Equilibrium points in n-person games. Proceedings of the National Academy of Sciences of the United States of America. 1950; 36 (1): 48–49, doi:10.1073/pnas.36.1.48.

Neumann   J von and Morgenster ON, Theory of Games and Economic Behavior. Princeton: P. Univ. Press; 2005.

Newman  ME, Strogatz SH, and Watts DJ, Random Graphs with Arbitrary Degree Distributions and Their Applications. Physical Review E. 2001; 64(2): 17.

Ohno S, Much "junk" DNA in our genome. In Evolution of Genetic Systems. H. H. Smith. ed. 1972; 366-370.

Ostrom E, A Behavioral Approach to Rational Choice Theory of Collective Action; Presidential Address, American Political Science Association, 1997.  The American Political Science Review. 92(1): 1-22; 1998.

Padilla-Walker L and Poole D, Memory of Previous Recall: A Comparison of Free and Cued Recall. Applied Cognitive Psychology. 2002; 6: 515-524.

Pattee, H.. H. (ed.),  Hierarchy theory: the challenge or complex systems.  New York: Braziller; 1973.

Palis J, Chaotic and Complex Systems. Curr Sci. 2002; 82(4): 403-406.

Pavlov IP, Conditioned Reflexes: An Investigation of the Physiological Activity of the Cerebral Cortex (translated by G. V. Anrep). London: Oxford University Press; 1927.

Pepper JW, Relatedness in Group-Structured Models of Social Evolution. Journal of Theoretical Biology. 2000; 206(3): 355-368.

Pepper JW and  Smuts BB, A Mechanism for the Evolution of Altruism Among Non-kin: Positive Assortment Through Environmental Feedback. American Naturalist. 2002; 160(2): 205-213.

Peterson JB and Carson S, Latent inhibition and openness to experience in a high-achieving student population. Personality and Individual Differences. 2000; 28: 323–332.

Perrie M, The Agrarian Policy of the Russian Socialist-Revolutionary Party from its Origins through the Revolution of 1905–07. Cambridge, UK: Cambridge University Press; 1976.

Raaijmakers JGW , `Retrieval from Long Term Store: A General Theory and Mathematical Models'  Unpublished Doctoral Dissertation, University of Nijmegen, The Netherlands; 1979.

Raaijmakers JGW and Shiffrin RM, `SAM: A Theory of Probabilistic Search in Associative Memory', In G. H. Bower (Ed.) The Psychology of Learning and Motivation: Advances in Research and Theory. New York: Academic Press. 1980: 207-262.

Ratiu P and  Talos IF, The tale of Phineas Gage. Digitally re-mastered.  N. Engl. J. Med. 2004; 351: e21-e21.

Rauch SL, Milad MR, Orr SP, Quinn BT, Fischl B, Pitman, R, Orbitofrontal thickness, retention of fear extinction, and extraversion. Neuroreport. 2005; 16(17):1909-12.

Radkey O, The Agrarian Foes of Bolshevism: Promise and Default of the Russian Socialist Revolutionaries, February to October 1917. Unpublished Doctoral Dissertation; 1958.

Reichardt J and  White D R, Role Models for Complex Networks.
European Physical Journal B. 2007; 60: 217-224.

Remnick D, Lenin's Tomb: The Last Days of the Soviet Empire. Vintage Books; 1994.

Sakwa R, The Rise and Fall of the Soviet Union, Routledge, London; 1999.

Schulkin J, The Neuroendocrine Regulation of Behavior,  NY: Cambridge Univ. Press; 1999.

Schwefel H,  Numerische Optimierung von Computer-Modellen (PhD thesis); 1974. Reprinted by Birkhäuser; 1977.

Shafee F, A Spin Glass Model of Human Logic Systems. xxx.lanl.gov. arXiv:nlin/0211013; 2002.

Shafee F, Chaos and Annealing in Social Networks. xxx.lanl.gov. arXiv:cond-mat/0401191; 2004.

Shafee F, Spin-glass-like Dynamics of Social Networks. xxx.lanl.gov.
arXiv:physics/0506161; 2005.

Shafee F, Oligo-parametric Hierarchical Structure of Complex System. NeuroQuantology J. 2007; 5(1):85-99.

Shafee F, Lambert function and a new non-extensive form of entropy. IMA Journal of Applied Mathematics. 2007b; 72: 785-800.

Shafee F,  Bifurcation of Identities and the Physics of Transition Economies. Transition Economies: 21st Century Issues and Challenges. ed Gergõ M. Lakatos. Nova Publishers; 2008.

Shafee F, Organization and Complexity in a Nested Hierarchical Spin-Glass like Social Space. submitted; 2009.

Shafee F, Agent Components and the Emergence of Altruism in Social Interaction Networks. xxx.lanl.gov.  arXiv:0901.3772v1; 2009b.

Shafee F, A Network of Perceptions.  Neuroquantology J. 2009c; (7):2.

Sherrington D   and Kirkpatrick SK, Solvable Model of Spin Glass. Phys. Rev. Lett.  1975; 35 :1792-1796.

Sinha S  and. Raghavendra S, Phase Transition and Pattern Formation in a Model of Collective Choice Dynamics. Santa Fe Paper. 04-09-028; 2004.

Tabeau, Ewa; Bijak J, War-related Deaths in the 1992–1995 Armed Conflicts in Bosnia and Herzegovina: A Critique of Previous Estimates and Recent Results.  European Journal of Population/ Revue europenne de Dmographie. 2005; 21: 187-215(29).

Tavares H,  Zilberman ML,  Hodgins DC, el-Guebaly N,
Comparison of Craving between Pathological Gamblers and Alcoholics.
Alcoholism: Clinical and Experimental Research. 2005; 29 (8): 1427–1431.

Trivers, RL, The evolution of reciprocal altruism. Quarterly Review of Biology. 1971; 46: 35-57.

United Nations High Commission for Refugees. `The humanitarian operation in Bosnia, 1992-95: the dilemmas of negotiating humanitarian access'; May 1999.

Vanhanen T, Domestic Ethnic Conflict and Ethnic Nepotism: A Comparative Analysis. Journal of Peace Research. 1999; 36(1): 55-73.

Weisbuch G,  Deffuant G,  Amblard F and  Nadal JP, Interacting Agents and Continuous Opinions Dynamics. Santa Fe Institute Working Paper.  01-11-072; 2001.

Winter, L and Uleman, JS, When are Social Judgements Made? Evidence of the Spontaneousness of Trait Interference. Journal of Personality and Social Psychology. 1984; 47: 237-252.

Wolfram S, A New Kind of Science. Wolfram Media, Inc; 2002.


\end{document}